\documentclass{aa}  

\usepackage{graphicx}

\usepackage{txfonts}
\begin{document}

   \title{H$_2$ chemistry in interstellar ices: The case of CO ice hydrogenation in UV irradiated CO:H$_2$ ice mixtures
   	}
\titlerunning{The surface chemistry of electronically excited species and H$_2$}
   \author{K.-J. Chuang\inst{1,2}, G. Fedoseev\inst{3}, D. Qasim\inst{1}, S. Ioppolo\inst{4, 5}, E.F. van Dishoeck\inst{2}, H. Linnartz\inst{1}
    }
\authorrunning{Chuang et al.}
   \institute{Sackler Laboratory for Astrophysics, Leiden Observatory, Leiden University, P.O. Box 9513, 
   	NL-2300 RA Leiden, the Netherlands
   	\\
              \email{chuang@strw.leidenuniv.nl}
              \and
              Leiden Observatory, Leiden University, P.O. Box 9513, NL-2300 RA Leiden, the Netherlands
              \and
              INAF – Osservatorio Astrofisico di Catania, via Santa Sofia 78, 95123 Catania, Italy
              \and
              School of Electronic Engineering and Computer Science, Queen Mary University of London, 
              Mile End Road, London E1 4NS, UK
              \and
              School of Physical Sciences, The Open University, Walton Hall, Milton Keynes MK7 6AA, UK
             }

  \date{July 2018}

 
  \abstract
   {In dense clouds, hydrogenation reactions on icy dust grains are key in the formation of molecules, like formaldehyde, methanol and complex organic molecules (COMs). These species form through the sequential hydrogenation of CO ice. Although molecular hydrogen (H$_2$) abundances can be four orders of magnitude higher than those of free H-atoms in dense clouds, H$_2$ surface chemistry has been largely ignored; several laboratory studies show that H$_2$ does not actively participate in \textquotedblleft non-energetic\textquotedblright~ice chemistry because of the high activation energies required.}
   {For the example of CO ice hydrogenation, we experimentally investigated the potential role of H$_2$ molecules on the surface chemistry when “energetic” processing (i.e., UV photolysis) is involved. We test whether additional hydrogenation pathways become available upon UV irradiation of a CO:H$_2$ ice mixture and whether this reaction mechanism also applies to other chemical systems.}
   {Ultra-high vacuum (UHV) experiments were performed at $8-20$ K. A pre-deposited solid mixture of CO:H$_2$ was irradiated with UV-photons. Reflection absorption infrared spectroscopy (RAIRS) was used as an in situ diagnostic tool. Single reaction steps and possible isotopic effects were studied by comparing results from CO:H$_2$ and CO:D$_2$ ice mixtures.}
   {After UV-irradiation of a CO:H$_2$ ice mixture, two photon-induced products, HCO and H$_2$CO, are unambiguously detected. The proposed reaction mechanism involves electronically excited CO in the following reaction steps: $\text{CO}+\text{h}\nu \longrightarrow \text{CO}^*$, $\text{CO}$$^*+\text{H}_2\longrightarrow \text{HCO}$+$\text{H}$
   where newly formed H-atoms are then available for further hydrogenation reactions. The HCO formation yields have a strong temperature dependence for the investigated regime, which is most likely linked to the H$_2$ sticking coefficient. Moreover, the derived formation cross section reflects a cumulative reaction rate that mainly determined by both the H-atom diffusion rate and initial concentration of H$_2$ at 8-20 K and that is largely determined by the H$_2$ sticking coefficient. Finally, the astronomical relevance of this photo-induced reaction channel is discussed.}
   {}


\maketitle{}
\section{Introduction}   
In dense molecular clouds, carbon monoxide starts heavily accreting on H$_2$O-rich ice mantles when densities (i.e., \textit{n}$_\text{H}$=$2$\textit{n}(H$_2$)+\textit{n}(H)) increase to $\sim$$10^{4-5}$ cm$^{-3}$, and temperatures drop to $\sim$10 K. This results in a CO-rich ice coating with a thickness of $\sim$$0.01$ $\mu$m \citep{Pontoppidan2006, Boogert2015} and is known as the \textquotedblleft CO catastrophic freeze-out stage\textquotedblright. During this stage, the simultaneous accretion of H-atoms and CO leads primarily to the formation of H$_2$CO and CH$_3$OH through successive H-atom addition reactions CO$\xrightarrow{\text{H}}$HCO$\xrightarrow{\text{H}}$H$_2$CO$\xrightarrow{\text{H}}$CH$_3$O$\xrightarrow{\text{H}}$CH$_3$OH as introduced in gas-grain models by \citet{Tielens1982}. These hydrogenation reactions have been investigated in a number of systematic laboratory experiments (\citealt{Watanabe2002, Fuchs2009}, see reviews by \citealt{Watanabe2008, Hama2013, Linnartz2015}) as well as astrochemical simulations and theoretical studies \citep{Charnley1997, Cuppen2009, Chang2012}. The CO+H channel is also regarded as a starting point in the formation of various complex organic molecules (COMs) in dense clouds; recombination of reactive intermediates \textemdash HCO, CH$_2$OH, and CH$_3$O, formed in H-atom addition and abstraction reactions with each other or with other reaction products results in the low temperature solid-state formation of larger COMs, like glycolaldehyde, ethylene glycol, glycerol, and likely glyceraldehyde \citep{Garrod2006, Woods2012, Butscher2015, Butscher2016, Fedoseev2015b, Fedoseev2017, Chuang2016, Chuang2017}. The astronomical gas-phase detection of a number of COMs in cold dark regions, that is environments in which (UV) photo-processing is not dominant, has been explained in this way \citep{Oberg2010, Bacmann2012, Cernicharo2012, Jimenez-Serra2016}, even though the mechanism transferring the solid COMs into the gas phase is still not fully understood \citep{Bertin2016, Chuang2018, Ligterink2018, Balucani2015}. Key in all this work is the important role of accreting H-atoms.

Molecular hydrogen, H$_2$, is the most abundant molecule in the Universe \citep{Wooden2004}. Particularly in molecular clouds, the gaseous abundance of H$_2$ is about four orders of magnitude higher than that of CO and H-atoms. Molecular hydrogen may freeze out and the direct observation of H$_2$ ice was claimed by \citet{Sandford1993b} in the infrared spectrum of WL5 in the $\rho$ Oph molecular cloud, thanks to a small induced dipole as a result of H$_2$ interacting with surrounding ice species \citep{Warren1980}. Later on this observation was questioned \citep{Kristensen2011}. It is, however, widely accepted that H$_2$ is abundantly formed in the solid state through H-H recombination on silicate or carbonaceous dust grains or amorphous water ice, through photolysis of hydrogenated amorphous carbon or water, or H$_2$ abstraction from polycyclic aromatic hydrocarbons (see reviews by \citealt{Vidali2013, Wakelam2017}, and references therein). Which of these formation mechanisms dominates, depends on the environmental conditions \citep{Wakelam2017}.

H$_2$ is extremely volatile, and its sticking coefficient, which is defined as the ratio of “adsorbed-species” to “total incident-species” for a given period of time, is a function of several properties, for example, binding energy of adsorbed species on a surface, species coverage, and substrate temperature \citep{Hama2013}. This coefficient also depends on the incident energy and angle of impacting species \citep{Matar2010}. Given the very small binding energy (E$_b$$\sim$100 K) of H$_2$ accreting on a preformed H$_2$ ice layer, multilayer “pure” H$_2$ ice is not expected under dense cloud conditions \citep{Lee1972}. However, in space, ice mantles formed by condensation on grain surfaces are typically amorphous, resulting in a distribution of binding energies. In a three-dimensional off-lattice Monte Carlo simulation, \citet{Garrod2013} found that the water ice formation on grain surfaces exhibits a highly porous (creviced) structure. Sites with relatively high binding energies in these pores can be a place where H$_2$ sticks \citep{Buch1994}. Laboratory results by \citet{Dissly1994} showed that H$_2$ can accrete together with other interstellar species, such as H$_2$O, under dense cloud conditions with an ice ratio (H$_2$:H$_2$O) of $\sim$0.3. However, a gas phase co-deposition of H$_2$O and H$_2$ in the interstellar medium (ISM) is unlikely due to the low density of gaseous H$_2$O for the temperatures at which H$_2$ starts freezing-out. Moreover, it is well established that H$_2$O is mainly produced in situ on grain surfaces through the hydrogenation of oxygen allotropes accreting from the gas phase prior to gaseous CO condensation \citep{Hiraoka1998, Dulieu2010, Miyauchi2008, Ioppolo2008, Ioppolo2010, Oba2009, Oba2012, Cuppen2010, Lamberts2013}. In recent laboratory studies, the reported H$_2$ sticking coefficients in a sub-monolayer regime on an olivine sample (i.e., a magnesium iron silicate) showed an unexpectedly high value of $\sim$0.7 at 10 K \citep{Acharyya2014}. This hints for the possibility that gaseous H$_2$ could co-deposit with other species on dust grains at low temperature and in particular with CO, forming CO:H$_2$ ice mantles, during its catastrophic freeze-out stage. In principle, this could facilitate reactions between CO and H$_2$, however, as ground-state molecule-molecule reactions typically have a very high activation energy and are often endothermic, even the four orders of magnitude higher H$_2$ abundance compared to atomic hydrogen is not expected to compensate for the high reaction barriers.

At high visual extinctions (A$_V$), dark clouds are shielded from external UV radiation by the dust. Only cosmic rays are expected to penetrate the cloud and react with the abundant gaseous H$_2$, resulting in the emission of UV photons. The internal cloud UV-photon flux is $(1-10)$$\times$$10^3$ photons cm$^{-2}$ s$^{-1}$, which is a few times lower than the typical H-atom flux in the same regions \citep{Prasad1983, Mennella2003, Shen2004}. However, UV-photons have a larger penetration depth in interstellar ice analogues, for instance, 830 ML in H$_2$O and 640 ML in CO, with 95\% absorption for 120-160 nm \citep{Cruz-Diaz2014a}. The energy of these UV-photons (mainly located around 160 and at $121.6$ nm, i.e., Ly$\alpha$) is not high enough to directly dissociate H$_2$ or CO, but is sufficient to radiatively pump CO (i.e., A$^1$$\Pi$$\longleftarrow$~X$^1$$\Sigma$$^+$) into a vibronically excited state, CO$^*$ \citep{Tobias1960}. This energy can be transferred from subsurface layers to eject CO molecules located in the first few top-layers resulting in a CO photo-desorption event, following a so called DIET (Desorption Induced by Electronic Transition) mechanism \citep{Fayolle2011, Bertin2013, vanHemert2015}. Alternatively, this energy can be used to overcome the activation energy of the involved barriers resulting in the formation of photo-products, such as CO$_2$ \citep{Gerakines1996, Gerakines2001, Loeffler2005}.

The present work is motivated by our limited understanding of the impact of UV-photons interacting with H$_2$-containing ices under dense cloud conditions. The idea of studying the role of molecular hydrogen in grain surface chemistry is not new. Previous work, like \citet{Fuchs2009} excluded the direct role of H$_2$ in the formation of H$_2$CO and CH$_3$OH. \citet{Oba2012} proposed a H$_2$O formation mechanism through the reactions between H$_2$ and OH radical, and found isotope effects when using D$_2$ instead of H$_2$ (see also \citealt{Meisner2017}). Recently, \citet{Lamberts2014} experimentally studied the reaction between H$_2$ and O in the ground state to form H$_2$O on grain surfaces, and this only resulted in a rather low upper limit for H$_2$O formation for this channel in dense clouds.  A similar mechanism was also studied to explain the HCN formation through the reaction H$_2$+CN in an H$_2$ matrix experiment \citep{Borget2017}. Here, we focus on the UV irradiation of a CO:H$_2$ ice mixture studied for temperatures ranging from $8$ to $20$ K. The aim is to investigate whether cosmic ray induced UV-photons can trigger surface reactions between an electronically excited species, for example, CO in this work, and H$_2$, the two most abundant molecules in prestellar cores, and how this compares to the regular CO ice H-atom addition reaction scheme. 
\section{EXPERIMENTAL}
\indent All experiments were performed by using SURFRESIDE$^2$, an ultra-high vacuum (UHV) setup, which has been described in detail in \citet{Ioppolo2013}. The base pressure of the main chamber is $\sim$10$^{-10}$ mbar, and the H$_2$O contamination from the residual gas accretion, which is observed by its monomer IR feature at $1600$ cm$^{-1}$, is estimated to be $<$$3$$\times$$10^{10}$ molecules cm$^{-2}$ s$^{-1}$. A gold-plated copper substrate is centered in the chamber and cooled by a closed-cycle helium cryostat that allows for variation of the substrate temperature between $8$ and $450$ K, which is monitored by two silicon diode thermal sensors with $0.5$ K absolute accuracy. Gaseous species, that are H$_2$ (D$_2$) (Linde 5.0) and CO (Linde 2.0), are separately introduced into the UHV chamber through the Hydrogen Atom-Beam Source line (filament is off and at room temperature) and a molecule dosing line, respectively.
\begin{figure*}[t!]
	\begin{center}
		\includegraphics[width=\textwidth]{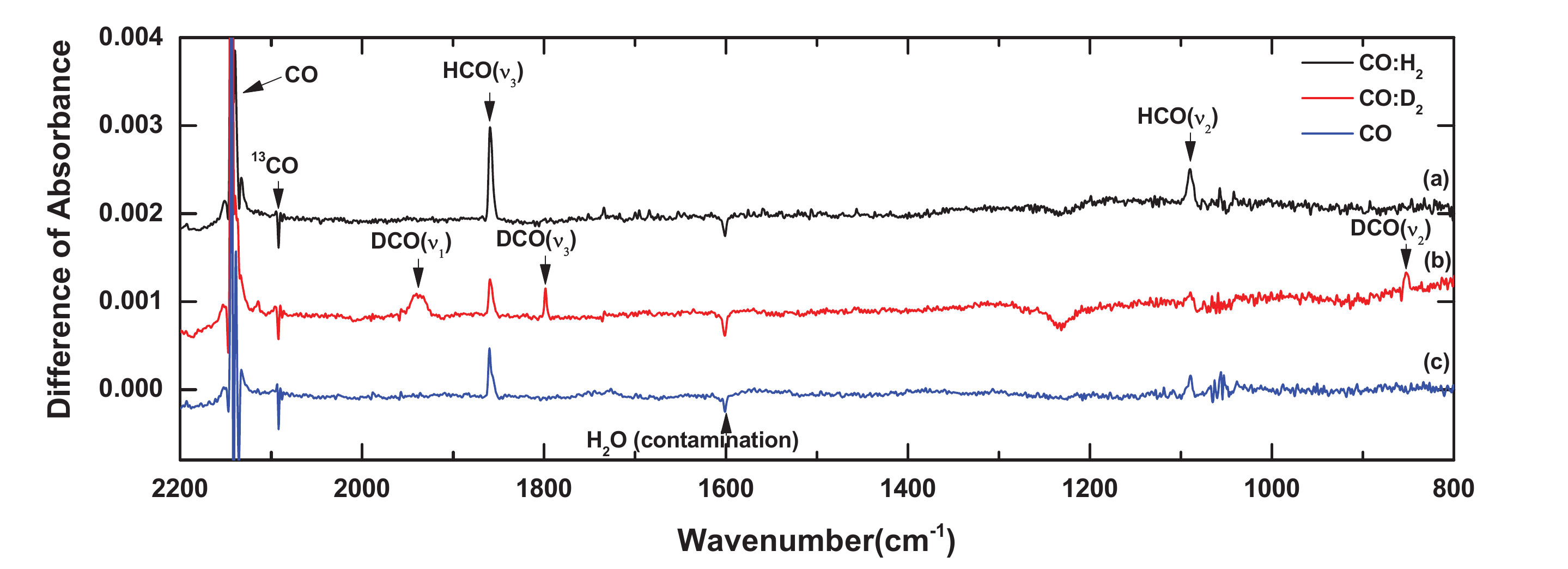}
		\caption{IR difference spectra obtained after UV irradiation of pre-deposited (a) CO:H$_2$, (b) CO:D$_2$, and (c) CO ice with a photon-flux of $6$$\times$$10^{12}$ photons cm$^{-2}$ s$^{-1}$ over $60$ minutes at $8$ K. 
		}
		\label{Fig1}
	\end{center}
\end{figure*}
The ice sample is monitored in situ, before and during UV-photon irradiation by Fourier Transform Reflection-Absorption InfraRed Spectroscopy (FT-RAIRS) in the range from $700$ to $4000$ cm$^{-1}$, with $1$ cm$^{-1}$ resolution. The RAIR band strength value of CO and possible formation products, like H$_2$CO is obtained from the laser interference experiments described by \citet{Chuang2018}. The HCO radical band strength value for RAIRS is calibrated from the averaged transmission value estimated by \citet{Bennett2007} and  \citet{Gerakines1996} multiplying a conversion factor (i.e., the ratio of \textquotedblleft reflection mode\textquotedblright~to \textquotedblleft transmission mode\textquotedblright~is $4.7$ for present IR setting). The conversion factor is based on the assumption that the band strength ratio of HCO to CO is constant in both transmission and reflection IR spectroscopy \citep{Oberg2009a}. Since the DCO band strength has not been reported in the literature, it is estimated by multiplying the HCO band strength value with a factor of $1.35$. This conversion factor is obtained under the assumption that the band strength ratio in C=O stretching mode of DCO to HCO is similar to the ratio of D$_2$CO to H$_2$CO (i.e., $1.35$) reported in \citet{Hidaka2009} . The used IR band strengths in this work are $5.2$$\times$$10^{-17}$, $5.7$$\times$$10^{-17}$, $7.7$$\times$$10^{-17}$, and $8.3$$\times$$10^{-17}$ cm molecule$^{-1}$ for CO ($2142$ cm$^{-1}$), HCO ($1859$ cm$^{-1}$), DCO ($1798$ cm$^{-1}$), and H$_2$CO ($1737$ cm$^{-1}$), respectively. The deposition rate of CO ice is $1.7$$\times$$10^{13}$ molecules cm$^{-2}$ s$^{-1}$ determined by using a modified Beer-Lambert law for the IR absorbance at $2142$ cm$^{-1}$. As H$_2$ (D$_2$) is an IR inactive molecule, the H$_2$ (D$_2$) exposure rate is obtained by the Langmuir estimation (i.e., $1$$\times$$10^6$ torr$\cdot$ s=$1$ L) resulting in a flux through the substrate plane of $7.3$$\times$$10^{14}$ molecules cm$^{-2}$ s$^{-1}$. The H$_2$ sticking coefficient (\textit{f}) on pure CO ice is not available from the literature, and is expected to depend strongly on temperatures in the range of $8-20$ K. Moreover, not all the H$_2$ that gets temporarily stuck will stay on the surface long enough to get also trapped in the bulk of the ice; the residence time of H$_2$ on the surface is limited which causes a fraction to return into the gas phase. Therefore, the estimated concentration of H$_2$ ice is an upper limit and expected to be less than CO due to the very small binding energy for multilayer H$_2$ ice. The ice deposition time is $60$ minutes resulting in a very similar column density of CO in all experiments.

After the simultaneous deposition of a CO:H$_2$ ice mixture, UV photons generated by a Microwave Discharge Hydrogen flowing Lamp (MDHL) are guided through a MgF$_2$ window onto the ice sample at a $90$ degrees angle with respect to the ice layer covering the entire substrate area ($2.5$$\times$$2.5$ cm$^2$). The spectral emission pattern and fluxes have been characterized in detail in previous work \citep{Ligterink2015b, Fedoseev2016, Chuang2017}. The used H$_2$ pressure amounts to $\sim$$1$ mbar which corresponds to a ratio of \textquotedblleft Ly$\alpha$\textquotedblright~to
\textquotedblleft H$_2$-emission ($\sim$$160$ nm)\textquotedblright$\cong$$1.7$ and UV-photon flux of $\sim$$6$$\times$$10^{12}$ photons cm$^{-2}$ s$^{-1}$ \citep{Ligterink2015b}. 
\section{RESULTS}
\textbf{HCO formation:} Figure \ref{Fig1} presents the IR difference spectra obtained after UV-photon irradiation of pre-deposited (a) CO:H$_2$, (b) CO:D$_2$ and (c) pure CO ice at $8$ K with a photon-flux of $6$$\times$10$^{12}$ photons cm$^{-2}$ s$^{-1}$ for $60$ minutes. The negative peaks visible at $2142$ and $2091$ cm$^{-1}$ are due to CO and its natural isotope $^{13}$CO, respectively, and reflect that their initial ice abundances decrease through photo-desorption and photo-chemistry. H$_2$ consumption cannot be monitored by using RAIRS; in the infrared, frozen H$_2$ can only be made visible through transitions at $4137$ and $4144$ cm$^{-1}$ that are the result of a small dipole moment induced through interactions with CO ice \citep{Warren1980}. The estimated band strength, however, is extremely small, that is $\sim$$10^{-19}$ cm molecule$^{-1}$ and about two orders of magnitude smaller than for CO ice \citep{Sandford1993a}. For the relatively thin ices studied in our experiments, therefore, it is not possible to monitor H$_2$ directly.

The positive peaks at $1859$, $1090$, and $2488$ cm$^{-1}$ (the latter is not shown in Figure \ref{Fig1}) in the CO:H$_2$ experiment indicate the formation of  HCO, and originate from its C-O stretching ($\nu$$_3$), bending ($\nu$$_2$), and C-H stretching ($\nu$$_1$) vibration mode, respectively \citep{Ewing1960, Milligan1964}. The corresponding isotopic product of the formyl radical (DCO) in the CO:D$_2$ experiment is identified by absorption signals at $1798$, $852$, and $1938$ cm$^{-1}$ due to its C-O stretching ($\nu$$_3$), bending ($\nu$$_2$), and C-D stretching ($\nu$$_1$) vibrational modes, respectively \citep{Ewing1960, Milligan1964}. In experiments (b) and (c), HCO peaks can also be observed. The HCO features in (b) and (c) are weaker than in (a) and can be explained by the presence of  H$_2$ or H$_2$O as background residual gases in the UHV chamber. This is confirmed by the fact that the HCO feature strengths are identical in experiments (b) and (c). Since no DCO signal is found in experiment (a) we confirm the formation of HCO(DCO) upon UV irradiation of a CO:H$_2$(CO:D$_2$) ices. The estimated H$_2$O column density, that is \textit{N}(H$_2$O), is below $\sim$$0.1$$\times$$10^{15}$ molecules cm$^{-2}$ in all three experiments and the ratio of \textit{N}(H$_2$O)/\textit{N}$_{\text{deposited}}$(CO)<$0.002$ after CO:H$_2$ ice mixture preparation. Other photolysis products, like CO$_2$ ($2346$ cm$^{-1}$) and its isotope $^{13}$CO$_2$ ($2280$ cm$^{-1}$) are also detected in all three experiments (not shown in Figure \ref{Fig1}). Their formation has been reported in previous UV-photon irradiation studies of pure CO ice \citep{Gerakines1995, Gerakines2001, MunozCaro2010, Chen2014, paardekooper2016b}.
\begin{figure}
	\begin{center}
		\includegraphics[width=90mm]{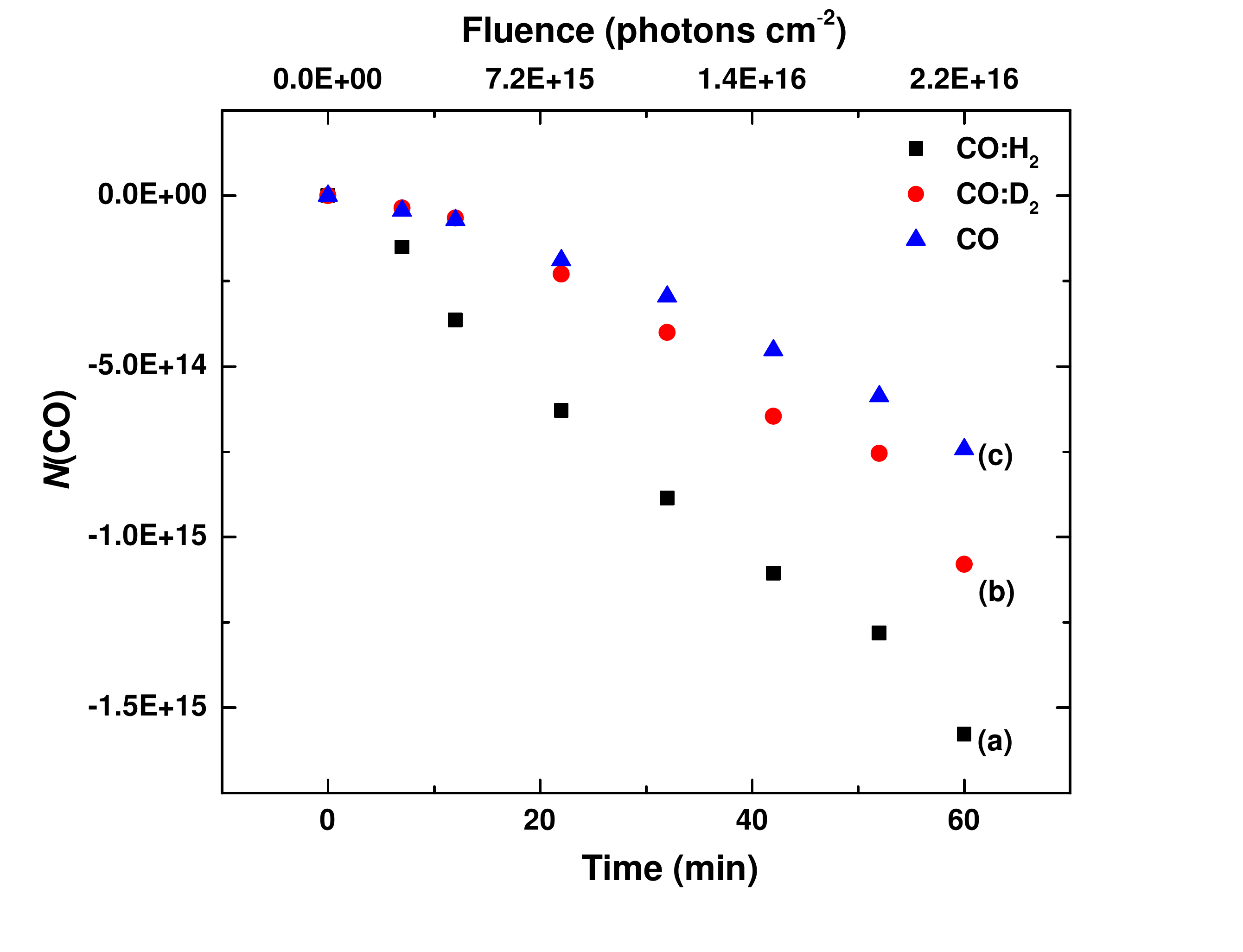}
		\caption{Evolution of the $\Delta$\textit{N}(CO) depletion over $60$ min of UV-photon irradiation with a photon-flux of $6$$\times$$10^{12}$ photons cm$^{-2}$ s$^{-1}$ at $8$ K for (a) CO:H$_2$, (b) CO:D$_2$, and (c) CO ice. The $\Delta$\textit{N}(CO) has been calibrated by the initial $^{13}$CO ice thickness.  
		}
		\label{Fig2}
	\end{center}
\end{figure}
The initial ice thickness of CO is beyond the RAIRS saturation limit for our experimental conditions, making it difficult to quantify the CO column density directly from its IR absorbance signal at $2142$ cm$^{-1}$. Given the constant CO:$^{13}$CO ratio (i.e., the natural abundance of CO:$^{13}$CO=$98.9$:$1.1$), the unsaturated $^{13}$CO feature can also be used to derive the CO abundance. Figure \ref{Fig2} shows the relative intensity changes of the CO abundance during $60$ minutes of UV-photon irradiation at $8$ K for the three experiments shown in Figure \ref{Fig1}, that is $\Delta$\textit{N}(CO), determined from the integrated IR absorbance area of $^{13}$CO ($2091$ cm$^{-1}$). In all three experiments the CO abundance is clearly decreasing; after $60$ minutes of UV photolysis which corresponds to a UV fluence of $2.2$$\times$$10^{16}$ photons cm$^{-2}$, the depletion $\Delta$\textit{N}(CO) is $1.6$$\times$$10^{15}$, $1.1$$\times$$10^{15}$, and $0.7$$\times$$10^{15}$ molecules cm$^{-2}$ for (a) CO:H$_2$, (b) CO:D$_2$, and (c) CO ice, respectively.

The UV-photon irradiation of pure CO ice has been extensively studied; CO ice is mainly found to photo-desorb following a DIET mechanism and small amounts of CO$_2$ at the level of a few percent are formed in the ice (see for an overview \citealt{paardekooper2016b}). After calibrating the distance between the ice sample and MDHL, and assuming that the light acts as a point source, the absolute depletion rate per second of pure CO ice is derived as $4.9$$\times$$10^{11}$ molecules cm$^{-2}$ s$^{-1}$, comparable to the rate of ($3.2$$\pm$$1.7$)$\times$$10^{11}$ molecules cm$^{-2}$ s$^{-1}$ reported in \citet{paardekooper2016b} for very similar MDHL settings. The derived absolute CO photo-desorption rate is $3.2$$\times$$10^{-2}$  molecules photon$^{-1}$ by employing the photon flux used in this work. The experimental CO absolute depletion rates for CO:H$_2$ and CO:D$_2$ ices are $\sim$$2$ and $\sim$$1.5$ times higher, respectively, indicating that more CO has been consumed than can be explained by photo-desorption only, namely, this observation is fully consistent with additional CO losses because of involvement in photo-chemical reactions with H$_2$ (D$_2$) forming HCO (DCO), as shown in Figure \ref{Fig1}.\\
\begin{figure}[]
	\begin{center}
		\includegraphics[width=90mm]{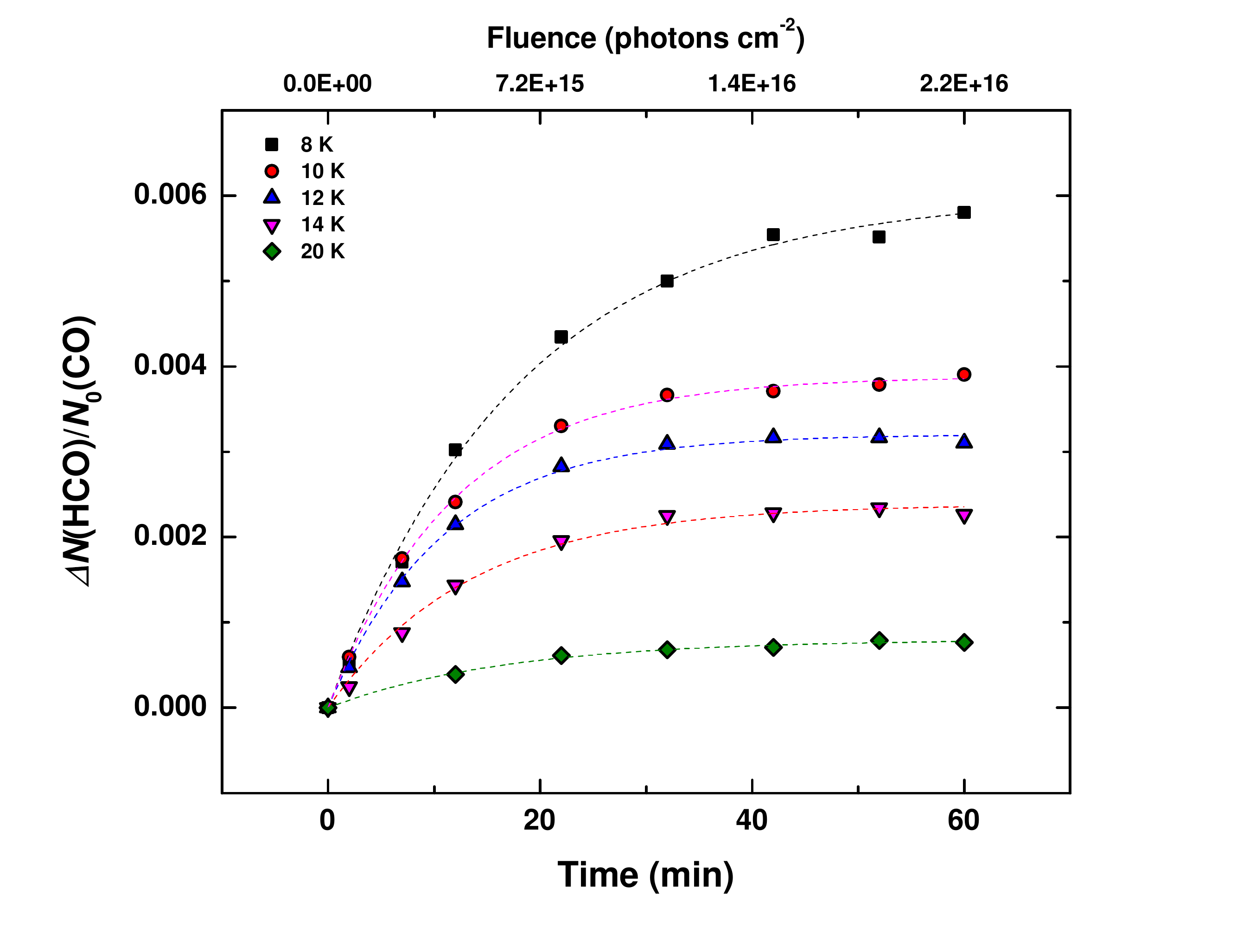}
		\caption{Evolution of the newly formed \textit{N}(HCO) w.r.t. initially predeposited \textit{N}$_0$(CO) over $60$ min of UV-photon irradiation of CO+H$_2$ ice mixture with a photon-flux of $6$$\times$$10^{12}$ photons cm$^{-2}$ s$^{-1}$ at $8$, $10$, $12$, $14$, and $20$ K, respectively. The dashed lines are the exponentially fitted results by using equation (\ref{Eq1}).  
		}
		\label{Fig3}
	\end{center}
\end{figure}
\\
\textbf{Temperature dependence:} Figure \ref{Fig3} presents the column density of \textit{N}(HCO) obtained from the integration of the IR feature at $1859$ cm$^{-1}$ and normalized to the initial CO abundance, that is \textit{N}$_0$(CO), for experiments of CO:H$_2$ ice over $60$ minutes UV-photon irradiation at $8$, $10$, $12$, $14$, and $20$ K. The HCO production grows with increasing UV-photon fluence impinging on the ice mixture, and reaches saturation around $30-60$ minutes (except at $8$ K), which implies that a balance between HCO formation and destruction processes is eventually reached. The final HCO abundances after $60$ min of UV-photon irradiation show a strong temperature dependence in the range of $8-20$ K; at $20$ K, the yield of HCO is only $\sim$10\% of the production at $8$ K. Because of the small absorption cross section reported for CO, that is $4.7$$\times$$10^{-18}$ cm$^2$ \citep{Cruz-Diaz2014a}, and as the energy dissipation lifetime is relatively short ($\sim$$15$ ns; \citealt{Chervenak1971}), here, we assume that the electronically excited CO triggered by impinging UV-photons is the limited reactant. Under the steady-state approximation the $\Delta$\textit{N}(HCO)/\textit{N}$_0$(CO) formation yield can be described by the following pseudo-first order kinetic equation \citep{Watanabe2006, Hidaka2007}:
\begin{equation}\label{Eq1}
\frac {\Delta\textit{N}(\text{HCO})}{N_0(\text{CO})}=\alpha(1-\text{exp}(-\textit{N}(\text{H$_2$})\cdot~k\cdot~t))=\alpha(1-\text{exp}(-\phi~\cdot~\sigma~\cdot~t)),
\end{equation}
where \textit{N}(H$_2$) is the column density of hydrogen molecules in molecules cm$^{-2}$, \textit{k} is the rate constant in cm$^{2}$ molecule$^{-1}$s$^{-1}$, $\alpha$ is the saturation value (unitless), $\phi$ is the UV-photon flux in photons cm$^{-2}$ s$^{-1}$, $\sigma$ is the effective formation cross section in cm$^2$ photon$^{-1}$, and \textit{t} is the experimental time in seconds. The rate constant (\textit{k}) cannot be derived here due to the IR-inactive H$_2$, but the formation cross section ($\sigma$) is available from experimental results \citep{Gerakines1996, Oberg2009a, Oba2018}. The derived fitting parameters are plotted in Figures \ref{Fig4} and \ref{Fig5}.
\begin{figure}
	\vspace{-6mm}
	\begin{center}
		\includegraphics[width=90mm]{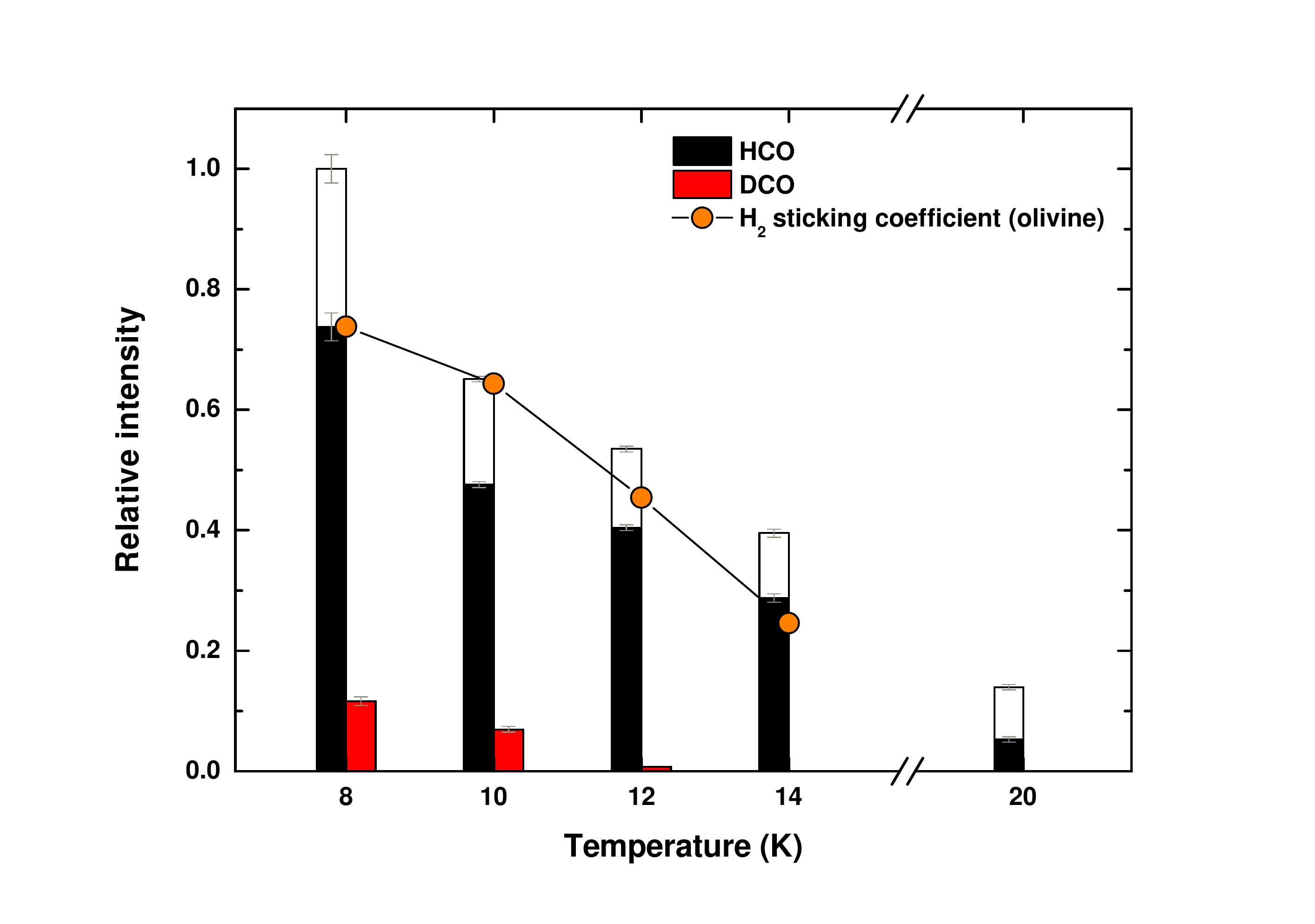}
		\caption{Relative intensity of the final yield for \textit{N}(HCO) (black column) and \textit{N}(DCO) (red column), i.e., the saturation value ($\alpha$), obtained after CO:H$_2$ and CO:D$_2$ ice UV irradiation for $60$ minutes with a flux of $6$$\times$$10^{12}$ photons cm$^{-2}$ s$^{-1}$ at temperatures ranging from $8$ to $20$ K, respectively. The white columns are the initial \textit{N}(HCO) before subtraction of the extra HCO contribution from the contamination of residual H$_2$ or H$_2$O in the UHV chamber. The column densities have been calibrated by the initial CO ice thickness, and further normalized to the maximum $\Delta$\textit{N}(HCO) in the CO:H$_2$ experiment at $8$ K, i.e., $2.5$$\times$$10^{14}$ molecules cm$^{-2}$. The orange points are the reported relative intensity of H$_2$ sticking coefficients on olivine substrate at $8-14$ K in \citet{Acharyya2014}.
		}
		\label{Fig4}
	\end{center}
\end{figure}
In Figure \ref{Fig4}, the relative intensity of the saturation value ($\alpha$), that is the final yield of HCO (black column) and DCO (red column), is presented for experiments of CO:H$_2$ and CO:D$_2$, respectively, after $60$ min of UV-photon irradiation for temperatures in the range from $8$ to $20$ K. The unwanted contributions of HCO (white column) from residual H$_2$ and H$_2$O gas in the UHV chamber (as shown in Figure \ref{Fig1}) can be independently quantified in the CO:D$_2$ experiments and subtracted from the final HCO production in the CO:H$_2$ experiments. The derived relative HCO abundance with respect to the maximum yield, namely, the total HCO abundance before subtracting the unwanted HCO contributions, obtained at $8$ K is $\sim$$0.74$, $\sim$$0.48$, $\sim$$0.40$, $\sim$$0.29$, and $\sim$$0.05$ at $8$, $10$, $12$, $14$, and $20$ K, respectively. The DCO abundance with respect to the HCO abundance at $8$ K is $\sim$$0.12$ and $\sim$$0.07$ at $8$ and $10$ K, respectively. At $12$ K, the DCO formation ratio, $\sim$$0.01$, is only regarded as an upper limit due to low S:N ratio from the corresponding RAIRS signal. 

In Figure \ref{Fig4}, the calibrated HCO abundances show a strong temperature dependence at $8-20$ K that can be linked to the relative sticking coefficient of H$_2$ reported in literature, for example, \citet{Acharyya2014} on olivine substrate. This is also shown in the figure. Clearly our results and the data points reported by \citet{Acharyya2014} are very similar. It should be noted, though, that the surface in our work, CO ice, is different from the substrate used in \citet{Acharyya2014}. As aforementioned, due to the lack of H$_2$ sticking coefficients on pure CO ice, the value reported for a non-water surface is currently the best option to compare with our experimental results. It reflects that the HCO formation abundance is predominantly controlled by the initial H$_2$ abundance in the pre-deposited CO:H$_2$ ice mixture. The H$_2$ abundance decreases with increasing temperature due to the dramatic drop of the sticking coefficient while the CO abundance (sticking coefficient=unity) remains constant at all experimental temperatures below $20$ K. Up to our knowledge, the D$_2$ sticking coefficient on any surfaces at temperatures in the range of $8-20$ K is not available from the literature. However, a similar correlation between the initial D$_2$ ice abundance and the final DCO yield is expected.

The formyl radical formation as result of UV-irradiation clearly shows an isotope effect in the overall formation yield; the HCO formation yield for UV irradiated CO:H$_2$ ice is about $6-7$ times higher than the corresponding amount of DCO in CO:D$_2$ ices. This observation is not directly in line with the assumption of a higher sticking coefficient for D$_2$ compared to H$_2$, namely, the initial amount of frozen D$_2$ for a specific temperature is expected to be higher than for H$_2$ due to the higher binding energies for D$_2$ \citep{Amiaud2015}. The observed difference in the formation yield between the CO:H$_2$ and CO:D$_2$ will be discussed in the next section.
\begin{figure}[b!]
	\begin{center}
		\includegraphics[width=90mm]{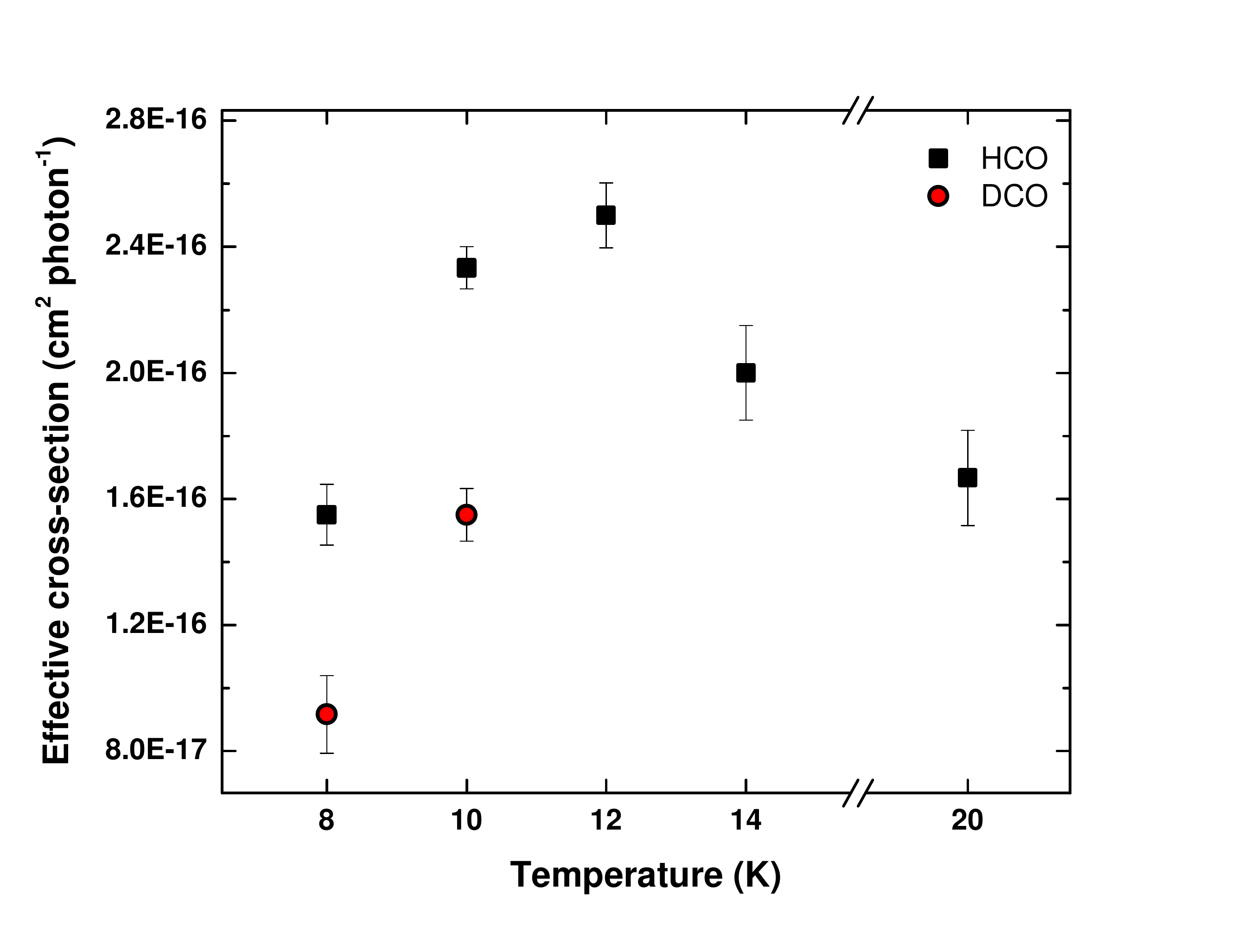}
		\caption{Derived effective formation cross section ($\sigma$) obtained by fitting the kinetics of \textit{N}(HCO) and \textit{N}(DCO) formation over $60$ minutes irradiation of UV-photons with a flux of $6$$\times$$10^{12}$ photons cm$^{-2}$ s$^{-1}$ in experiments CO:H$_2$ and CO:D$_2$ at temperatures in the range from $8$ to $20$ K, respectively.
		}
		\label{Fig5}
	\end{center}
\end{figure}
Figure \ref{Fig5} presents the effective formation cross section ($\sigma$), which is derived by the single exponential fit (i.e., equation (\ref{Eq1})) of HCO and DCO formation kinetics for product HCO (black squares) and DCO (red dots). The derived cross section is the cumulative outcome of multiple reaction channels of HCO (DCO), and is believed to be mainly controlled by two reaction parameters, such as diffusion rate and sticking coefficient (initial ice abundance). Figure \ref{Fig5} shows a temperature dependency of the HCO formation rate in the range of $8-20$ K, with a maximum value at $\sim$$12$ K. A proportional correlation, namely, increasing the ice sample temperature results in an increased formation cross section, in the range of $8-12$ K; at $12$ K the effective cross section of HCO is $1.7$ times higher than the value at $8$ K. One of the possible interpretations is that this positive slope can be explained by the diffusion rate that is expected to be also temperature dependent, namely, the higher the temperature, the higher the diffusion rate of H(D)-atoms is \citep{Fuchs2009}. However, at the same time, the formation cross section is also dominated by the available H$_2$ (D$_2$) ice abundance (species concentration in bulk ice) that is controlled by its sticking coefficient. At higher temperature, the concentration of H$_2$ (D$_2$) molecules in the CO ice is lower than that at low temperature. This results in a decreased effective cross section with increasing temperature from $2.5$$\times$$10^{-16}$ at $12$ K to $1.7$$\times$$10^{-16}$ cm$^{2}$ photon$^{-1}$ at $20$ K (i.e., $68$\% less). A detailed temperature dependent study (e.g., a wider temperature range and a better temperature resolution) is needed for a more detailed picture. For DCO, as aforementioned, the absolute formation yield above $12$ K cannot be derived from the IR spectrum due to the limited detection sensitivity. With much thicker ice than used in this work, the effective formation cross section for higher temperature can be measured and also for CO:D$_2$ ices but this is presently outside the scope of this work. A similar temperature dependence of the formation rate due to the different sticking coefficients was reported in previous laboratory studies of CO hydrogenation in the same temperature range of $8-20$ K \citep{Watanabe2006}.\\
\begin{figure}
	\begin{center}
		\includegraphics[width=90mm]{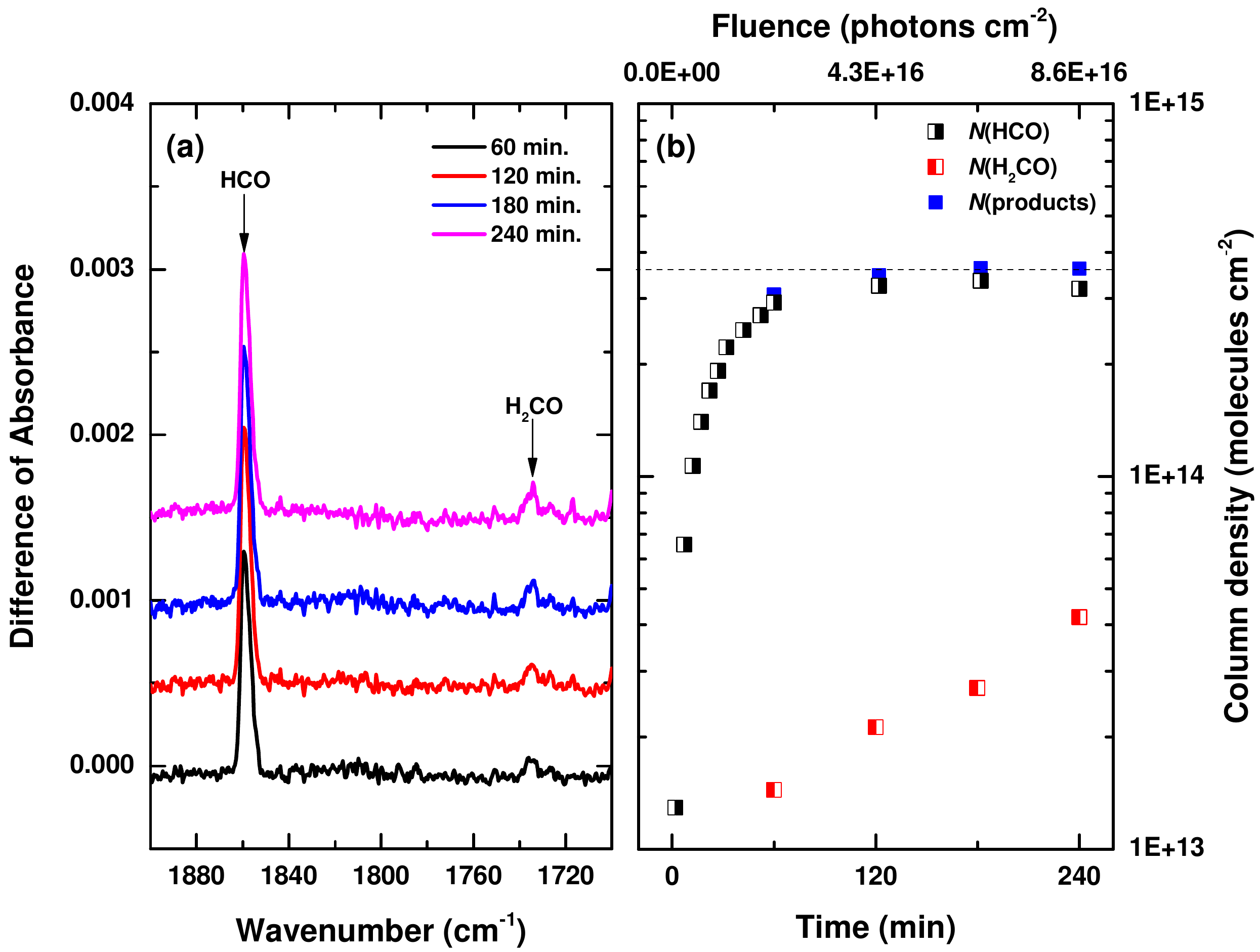}
		\caption{Left: IR difference spectra of the UV photolysis of pre-deposited CO:H$_2$ ice with a photon-flux of $6$$\times$$10^{12}$ photons cm$^{-2}$ s$^{-1}$ for $240$ min at $8$ K. Right: evolution of newly formed \textit{N}(HCO), \textit{N}(H$_2$CO) and total abundance of photo-induced products over $240$ min of UV-photon irradiation.
		}
		\label{Fig6}
	\end{center}
\end{figure}
\\
\textbf{H$_2$CO formation:} Figure \ref{Fig6}(a) shows the IR difference spectra obtained after UV irradiation of pre-deposited CO:H$_2$ for $240$ min with a flux of $6$$\times$$10^{12}$ photons cm$^{-2}$ s$^{-1}$ at $8$ K. As aforementioned, HCO is observed at $1859$ cm$^{-1}$ along with $2488$ and $1090$ cm$^{-1}$, and its abundance is relatively stable in the range from $60$ to $240$ min. The new IR feature appearing at $1737$ cm$^{-1}$ can be assigned to monomer H$_2$CO ($\nu$$_2$), and becomes much clearer with longer UV-photon irradiation time \citep{Khoshkhoo1973, Nelander1980}. In Figure \ref{Fig6}(b), the product abundances obtained from the integration of HCO ($1859$ cm$^{-1}$) and H$_2$CO ($1737$ cm$^{-1}$) IR signals, respectively, are presented over $240$ min. The HCO formation rate starts slowing down after $60$ minutes and slightly decreases after passing the maximum column density of $3$$\times$$10^{14}$ molecules cm$^{-2}$. The detectable abundance of H$_2$CO builds up after $60$ minutes of UV-irradiation as the fluence amounts to $2.2$$\times$$10^{16}$ photons cm$^{-2}$, and the final abundance ratio of $\Delta$\textit{N}(H$_2$CO)$/$$\Delta$\textit{N}(HCO) is $\sim$$0.13$ reflecting that H$_2$CO is a second generation product after HCO. The sum of \textit{N}(HCO) and \textit{N}(H$_2$CO) shows a relatively constant value of $3.6$$\times$$10^{14}$ after $120$ minutes suggesting a direct chemical link between these two products.
\section{Discussion}
\indent The UV-photon energy applied here is $\leq$$10.2$ eV, and cannot directly dissociate CO or H$_2$ molecules; the required threshold dissociation energies of CO and H$_2$ in the gas phase are $11.09$ and $11.20$ eV, respectively \citep{Field1966, Dalgarno1970, Okabe1978}. In the solid state these values can decrease, but with $0.13$ eV decrease for CO the effect is negligible \citep{Lu2005}. However, CO has a strong absorption cross section in the $127-157$ nm range, coinciding with the wavelength of the impacting UV-photons. This allows to excite CO into its first electronic state (A$^1$$\Pi$$\longleftarrow$X$^1$$\Sigma$$^+$), namely, CO$^*$ with excess energy $\geq$$7.9$ eV \citep{Lu2005, Mason2006, Cruz-Diaz2014a};
\begin{equation}\label{Eq2}
\text{CO}+\text{h}\nu\longrightarrow \text{CO}^*.
\end{equation}
Subsurface CO$^*$ can transfer the energy to the top layer molecules leading to the CO non-thermal desorption at low temperature, as discussed by \citet{Fayolle2011} and \citet{vanHemert2015}, and in the bulk it can further react with neighboring CO to form CO$_2$ \citep{Bertin2013, Okabe1978, Gerakines1996, Gerakines2001, Loeffler2005, Chen2014};
\begin{equation}\label{Eq3}
\text{CO}^*+\text{CO}\longrightarrow \text{CO}_2+\text{C}.
\end{equation}
Due to a small amount of H$_2$O contamination as shown in our IR spectrum at $1600$ cm$^{-1}$, another reported CO$_2$ formation channel through the reaction between CO and OH radical, originating from H$_2$O photodissociation, cannot be excluded \citep{Watanabe2002, Watanabe2007, Oba2010, Ioppolo2011}. 

As mentioned, the H$_2$ (D$_2$) molecule cannot be directly dissociated in H-atoms (D-atoms) upon Ly$\alpha$ irradiation, or electronically excited by an UV-photon ($\leq$$10.2$ eV), however, an alternative channel may apply that is similar to the CO$_2$ formation scheme shown in equation (\ref{Eq3}):
\begin{equation}\label{Eq4}
\text{CO}^*+\text{H}_2 \longrightarrow \text{HCO} +\text{H}_{\text{dis}}.
\end{equation}
The lowest potential energy of the first electronic state of CO is $7.9$ eV, and is larger than the reported enthalpy of CO+H$_2$$\longrightarrow$HCO+H in the gas phase, that is $\sim$$3.7$ eV \citep{Reilly1978}. Moreover, ab initio calculations show that the H$_2$ dissociation due to energy transfer from the electronically excited CO is a barrierless reaction \citep{Sperlein1987}. Therefore, we expect that the reaction shown in equation (\ref{Eq4}) proceeds without activation barrier in the solid state as well. The product HCO is likely thermally stabilized below $20$ K, and preserved in the ice mixture. The free H-atom (H$_{\text{dis}}$) that is formed in equation (\ref{Eq4}) may react with the newly formed HCO radical in equation (\ref{Eq4}) yielding H$_2$CO or CO+H$_2$ through:
\begin{subequations}\label{Eq5}
\begin{equation}\label{Eq5a}
\text{H}_{\text{dis}}+\text{HCO}\longrightarrow \text{H$_2$CO} 
\end{equation}
\begin{equation}\label{Eq5b}
\text{H}_{\text{dis}}+\text{HCO}\longrightarrow \text{CO}+\text{H$_2$}.
\end{equation}
\end{subequations}
There is also a possibility that the H$_{\text{dis}}$ diffuses in the bulk ice and reacts with CO to form HCO through direct H-atom addition reactions contributing to the total yield of HCO detected in the IR spectrum \citep{Watanabe2002, Fuchs2009}:
\begin{equation}\label{Eq6}
\text{H}_{\text{dis}}+\text{CO}\longrightarrow \text{HCO}
\end{equation}
or meets with other free radicals available in the ice, like HCO and H$_{\text{dis}}$ forming H$_2$CO (equation \ref{Eq5a}), H$_2$+CO (equation \ref{Eq5b}) and H$_2$ through:
\begin{equation}\label{Eq7}
\text{H}_{\text{dis}}+\text{H}_{\text{dis}}\longrightarrow \text{H}_2.
\end{equation}
The reaction channels of equations (\ref{Eq5}), (\ref{Eq6}), and (\ref{Eq7}) compete with each other showing a strong temperature dependence due to the different diffusion rates; at higher ice temperatures, the mobile H-atom will diffuse away much quicker from the newly formed HCO in the bulk ice, and eventually react with CO forming HCO. This may explain the increase of the observed HCO formation cross section in Figure \ref{Fig5} at temperatures from $8$ to $12$ K. Beyond $12$ K, the substantial decrease of the effective concentration of H$_2$ leads to a lower cumulative formation rate.

For the final yield of HCO, it is important to note that the HCO formation in equation (\ref{Eq4}) induced by UV-photons is expected to proceed without activation barrier, but the hydrogenation of CO in equation (\ref{Eq6}) was shown before to exhibit a strong isotope effect due to the quantum tunneling of H(D)-atoms \citep{Hidaka2007}. This likely explains the difference in formation yield between HCO and DCO, and shows that secondary D(H)-atoms are indeed contributing to the overall abundances.  Therefore, the derived formation cross section of formyl radical is concluded to be an effective value that is controlled by the reaction parameters, such as sticking coefficient, diffusion rate for H$_2$ and D$_2$ molecules, and an isotope effect that gets important in a secondary reaction step..

The possible formation mechanism of H$_2$CO is through the reaction HCO+H, of which the H-atom can be directly formed as shown in equation (\ref{Eq4}) or be produced by photodissociation of HCO \citep{Heays2017};
\begin{equation}\label{Eq8}
\text{HCO}+\text{h}\nu\longrightarrow \text{CO}+\text{H}.
\end{equation}
Alternatively, the interaction between two HCO radicals can result in their recombination into glyoxal or H$_2$CO and CO formation through the reaction \citep{Butscher2017}:
\begin{equation}\label{Eq9}
\text{HCO}+\text{HCO}\longrightarrow \text{H$_2$CO}+\text{CO}.
\end{equation}
In this work, the observed H$_2$CO formation is found when HCO abundance passes its maximum yield and starts decreasing (Figure \ref{Fig6}), implying that the HCO is a key precursor to form H$_2$CO in present study.

We note that the interaction of UV-photons and a metallic substrate (gold; work function=4.4 eV) may result in photoelectrons. These are expected to have a limited penetration depth in condensed ices, namely, few layers at maximum \citep{Jo1991}. To proof this assumption, a control experiment with a thick layer of Ar ice ($\sim$20 Langmuir) between gold substrate and CO:H$_2$ ice mixture has been performed. After applying the same fluence of UV-photon irradiation, the same absolute production yield of HCO is found; this implies that indeed any additional contribution from photo induced electrons interacting with the CO:H$_2$ ice mixture will be negligible.
\section{Astrochemical implication}
This laboratory work shows that electronically excited ice species induced by UV-photons can react with H$_2$ molecules adsorbed (or trapped) in interstellar ices at low temperatures to form new species. In space, cosmic rays, electrons, and photons can all interact with the ice mantle resulting in energy transfer. Such events can lead to a series of different thermal and chemical processes depending on the energetic input and chemical composition of the ice. For instance, previous laboratory studies on UV-irradiation of CO ice showed that although UV-photons cannot photodissociate CO molecules, they can electronically excite CO$^*$ species that can then cause the non-thermal desorption of other CO molecules \citep{Fayolle2011, vanHemert2015}. This mechanism successfully explains the observed gaseous abundance of CO at low temperature and also, at least partially, the photo-excitation chemistry resulting in the formation of C$_n$O$_m$ species in the solid phase, particularly CO$_2$ \citep{Loeffler2005, Gerakines2001}. It also has been linked to the location of photo-induced snow lines in proto-planetary disks \citep{Qi2015, Oberg2015}.
 
In dense clouds, the H$_2$ abundance is four orders of magnitude higher than that of H-atoms. Therefore, since H$_2$ molecules are expected to be found in interstellar ices at low temperatures \citep{Sandford1993b, Buch1994, Dissly1994}, it is possible that electronically excited species react with molecular hydrogen getting hydrogenated. The decomposition of H$_2$ induced by the excited species results in a free H-atom that can further hydrogenate other molecules forming hydrogen-saturated species, like CH$_3$OH and COMs. This mechanism may also apply to other reaction chains. For instance, water and hydrocarbons can be formed through reactions of hydrogen molecules with
photo-excited oxygen-atoms (O$^*$) and carbon-atoms (C$^*$), respectively. Both have a strong UV-photon absorption cross section in the range of $120-160$ nm. 

It should be noted that the solid state formation of hydrogen-rich species containing atomic C and O (e.g., H$_2$CO, and CH$_3$OH) is believed to occur predominantly through H-atom addition reactions to CO in dense clouds. In the present work we show for a first time that electronically excited species can react with frozen H$_2$ under dense cloud conditions, a process that is limitedly involved in current astrochemical networks. We propose a general mechanism on the example of CO:H$_2$+$\text{h}\nu$ as this system is well studied and has a minimum number of competing side-photochemical products. It offers an additional channel that holds the potential to form HCO radicals (and larger species) in interstellar ices, especially at very low temperatures. Since there is no proof of CH$_3$OH formation in our experiments, this hints at a lower efficiency of the CO$^*$+H$_2$ compared to the regular CO+H hydrogenation chain. As UV light can penetrate deeper in ices than H-atoms, the mechanism studied here may increase the abundance of HCO radicals in the deeper layers of the ice, where CO and H$_2$ (depending on temperature) are preserved and H-atoms cannot penetrate. Ultimately, this may affect the efficiency with which COMs can be formed in the later stage when the ice mantle is gently heated by the central protostar. The work presented here should be considered as a case study, investigating a new process capable of triggering chemical reactions at low temperatures in interstellar ices. The process may be more generally relevant, and applicable to other reaction chains as well. To which extent such reaction routes contribute to the full chemical picture, will be topic of future astrochemical modeling studies. 

\section{Conclusions}
Below the main findings of this experimental study are given for frozen H$_2$ when adsorbed or trapped in interstellar ice analogues upon UV irradiation: 
   \begin{enumerate}
      \item	UV-photons generated by cosmic rays in dense clouds may increase HCO (H$_2$CO and possibly larger COMs) abundances by triggering a solid-state reaction involving electronically excited CO$^*$ and H$_2$. The HCO formation is explained by two consecutive reaction steps CO$^*$+H$_2$$\longrightarrow$HCO+H$_{\text{dis}}$ and CO+H$_{\text{dis}}$$\longrightarrow$HCO.
      \item	The derived effective formation cross section shows a temperature dependence that is determined by the cumulative effect of the H-atom diffusion rate and initial H$_2$ concentration in the bulk ice that is determined by the sticking coefficient. For the investigated laboratory settings, we find a maximum at $12$ K of $2.5$$\times$$10^{-16}$ cm$^2$ photon$^{-1}$.      
      \item	The surface temperature between $8$ and $20$ K determines the photolysis product yield and this is due to the temperature dependent sticking efficiency of H$_2$. 
      \item	The mechanism that involves reactions between electronically excited species and H$_2$ molecules on icy grains may be of more general importance in ISM chemistry. It will take more detailed astrochemical modeling to put this pathway into perspective and to check to which extent this contributes to the formation of H$_2$CO and eventual COMs in space. Based on the experiments performed here, the mechanism is found to contribute, but very likely at a level that is (substantially) lower than regular CO hydrogenation upon impacting H-atoms.\\ 
   \end{enumerate}

\begin{acknowledgements}
This research was funded through a VICI grant of NWO (the Netherlands Organization for Scientific Research), an A-ERC grant 291141 CHEMPLAN, and has been performed within the framework of the Dutch Astrochemistry Network. Financial support by NOVA (the Netherlands Research School for Astronomy) and the Royal Netherlands Academy of Arts and Sciences (KNAW) through a professor prize is acknowledged. G.F. acknowledges the financial support from the European Union’s Horizon 2020 research and innovation program under the Marie Sklodowska-Curie grant agreement no. 664931. S.I. acknowledges the Royal Society for financial support and the Holland Research School for Molecular Chemistry (HRSMC) for a travel grant. The described work has benefited a lot from continuing collaborations within the framework of the FP7 ITN LASSIE consortium (GA238258) We thank M. van Hemert and T. Lamberts for stimulating discussions. 
\end{acknowledgements}

\bibliography{Ref}{}
\bibliographystyle{aa}

\end{document}